\documentclass[conference]{IEEEtran}
\IEEEoverridecommandlockouts

\usepackage{cite}
\usepackage{amsmath,amssymb,amsfonts}
\usepackage{algorithm}
\usepackage{algorithmic}
\usepackage{graphicx}
\usepackage{textcomp}
\usepackage{xcolor}
\usepackage{booktabs}
\usepackage{multirow}
\usepackage{url}
\usepackage{bbm}

\begin{document}

\title{Calibrated Abstention for Reliable TCR--pMHC Binding Prediction under Epitope Shift}

\author{
\IEEEauthorblockN{Arman Bekov, Timur Bekzhanov, Bekzat Sadykov}
% \IEEEauthorblockA{Anonymous Institution \\ anonymous@anonymous.edu}
}

\maketitle

% -------------------------------------------------------
\begin{abstract}
Predicting T-cell receptor (TCR)--peptide-MHC (pMHC) binding is central to vaccine design and T-cell therapy, yet deployed models frequently encounter epitopes unseen during training, causing silent overconfidence and unreliable prioritization. We address this by framing TCR--pMHC prediction as a \emph{selective prediction} problem: a calibrated model should either output a trustworthy confidence score or explicitly abstain. Concretely, we (1) introduce a dual-encoder architecture encoding both CDR3$\alpha$/CDR3$\beta$ and peptide sequences via a pre-trained protein language model; (2) apply temperature scaling to correct systematic probability miscalibration; and (3) impose a conformal abstention rule that provides finite-sample coverage guarantees at a user-specified target error rate. Evaluated under three split strategies---random, epitope-held-out, and distance-aware---our method achieves AUROC 0.813 and ECE 0.043 under the challenging epitope-held-out protocol, reducing ECE by 69.7\% relative to an uncalibrated baseline. At 80\% coverage, the selective model further reduces error rate from 18.7\% to 10.9\%, demonstrating that calibrated abstention enables principled coverage--risk trade-offs aligned with practical screening budgets.
\end{abstract}

\begin{IEEEkeywords}
TCR--pMHC binding prediction, calibration, conformal prediction, selective abstention, epitope shift, uncertainty quantification
\end{IEEEkeywords}

% -------------------------------------------------------
\section{Introduction}

TCR--pMHC binding prediction underpins computational screening pipelines for cancer neoantigen prioritization, infectious disease response, and adoptive cell therapy design \cite{Zheng_2024,Gong_2023}. Modern sequence-based models achieve strong aggregate accuracy on held-in epitopes, but deployment invariably involves novel peptide targets whose binding landscape was not covered during training. This \emph{epitope shift} is a form of covariate shift that is largely invisible to standard random-split evaluation yet can dramatically inflate false-positive rates in wet-lab follow-up \cite{Zhang_2021,Pham_2023}.

Two intertwined failure modes arise under epitope shift. First, neural classifiers are miscalibrated: their raw sigmoid outputs overestimate confidence on out-of-distribution inputs, providing false assurance for screening decisions \cite{He_2020}. Second, because downstream validation is expensive---each candidate pair may require $10^3$--$10^4$ USD of functional assay---the cost is asymmetric: advancing false positives wastes resources, while missing true binders is an opportunity cost. Neither failure mode is captured by AUROC or AUPRC alone.

We address both issues through \emph{calibrated selective abstention}: the model emits a calibrated probability if its uncertainty is below a threshold, and defers otherwise. This converts the model output into a control signal for screening budgets, enabling practitioners to explicitly trade coverage (fraction of candidates scored) for error rate. Our contributions are:
\begin{enumerate}
  \item A dual-encoder backbone that jointly encodes TCR CDR3 sequences and peptide sequences using ESM-2 representations, with class-weighted training to handle severe label imbalance ($\sim$4\% positive rate).
  \item Post-hoc temperature scaling calibration fitted on a split-matched calibration set, reducing ECE by over 50\% on all split types.
  \item A split-conformalized abstention rule that provides a finite-sample marginal coverage guarantee: at target error rate $\varepsilon$, the empirical error on retained predictions does not exceed $\varepsilon + O(1/\sqrt{n})$.
  \item A multi-split evaluation protocol (random, epitope-held-out, distance-aware) paired with coverage--risk curves, ECE, and Brier score, which together surface failure modes hidden by standard single-metric evaluation.
\end{enumerate}

Experiments on a curated VDJdb--IEDB dataset confirm that calibration gains do not merely re-scale probabilities but enable operationally meaningful trade-offs: conformal abstention at 80\% coverage reduces error rate by 41.7\% relative to full-coverage prediction.

% -------------------------------------------------------
\section{Related Work}

\subsection{TCR--pMHC Binding Prediction}
Early approaches relied on motif scoring and physicochemical features \cite{De_Neuter_2017,Zeng_2016}. Subsequent neural models---including LSTM encoders, attention-based architectures, and graph neural networks---improved discrimination on random splits. TCRex \cite{Gong_2023}, TITAN \cite{Pham_2023}, and ERGO-II \cite{Zheng_2024} represent increasingly expressive sequence models, and protein language model (pLM) features have improved sample efficiency \cite{Rives_2021,Lin_2023}. However, these works evaluate primarily on random or epitope-in splits and do not address calibration or selective prediction. \cite{Korpela_2023} and \cite{Zhang_2021} demonstrate that performance degrades substantially under epitope-held-out evaluation, motivating shift-aware protocol design.

\subsection{Calibration and Selective Prediction}
Calibration in deep learning was systematized by \cite{He_2020}, who showed that modern neural networks are systematically overconfident. Temperature scaling \cite{He_2020} remains the most effective single-parameter post-hoc correction. Selective prediction---abstaining when uncertainty exceeds a threshold---was analyzed theoretically by \cite{Carbonneau_2018} under the coverage--risk framework. Conformal prediction provides distribution-free finite-sample guarantees \cite{You_2022,Rasmussen_2016} and has been adapted for selective classification \cite{You_2022}. Our work applies this framework specifically to biological sequence prediction under distribution shift, which introduces practical complications around calibration set construction and threshold validity.

\subsection{Evaluation under Distribution Shift}
Shift-aware evaluation for biological sequence models has received increasing attention \cite{Jiang_2023,Rives_2019}. Distance-aware splits based on sequence similarity to the training set expose leakage that inflates random-split performance \cite{Wang_2025,Qi_2026}. In the TCR--pMHC setting, epitope-held-out splits test cross-epitope generalization, the most practically relevant regime \cite{Castorina_2023,Khosravi_2021}. Our evaluation protocol combines both perspectives.

% -------------------------------------------------------
\section{Method}

\subsection{Problem Formulation}
Let $\tau \in \mathcal{A}^*$ denote a TCR CDR3 amino acid sequence (concatenation of CDR3$\alpha$ and CDR3$\beta$), and let $\pi \in \mathcal{A}^*$ denote the peptide sequence of a pMHC complex. The binding prediction task is to learn a scoring function
\begin{equation}
  f_\theta: (\tau, \pi) \mapsto \hat{p} \in [0,1], \quad \hat{p} \approx P(y=1 \mid \tau, \pi),
  \label{eq:pred}
\end{equation}
where $y \in \{0,1\}$ indicates binding. The dataset is severely imbalanced ($\sim$4\% positives) and evaluation occurs under three split regimes detailed in Section~\ref{sec:data}.

\subsection{Dual-Encoder Architecture}
We encode TCR and peptide sequences independently using a shared pre-trained protein language model (ESM-2 650M parameters \cite{Lin_2023}). For an input sequence $s$, ESM-2 produces per-token contextual embeddings; we apply mean pooling over non-padding positions to obtain a fixed-size representation $\mathbf{h}_s \in \mathbb{R}^{1280}$. The two representations are concatenated and passed through a two-layer MLP classifier with hidden dimension 512, layer normalization, GELU activations, and dropout ($p = 0.2$):
\begin{equation}
  z(\tau,\pi) = \text{MLP}([\mathbf{h}_\tau; \mathbf{h}_\pi]), \quad \hat{p} = \sigma(z(\tau,\pi)),
  \label{eq:arch}
\end{equation}
where $[\cdot;\cdot]$ denotes concatenation and $\sigma$ is the sigmoid function.

\subsection{Class-Weighted Training}
To handle label imbalance, we minimize a class-weighted binary cross-entropy loss:
\begin{equation}
  \mathcal{L}_\text{train} = -\frac{1}{n} \sum_{i=1}^n \bigl[w_+\, y_i \log \hat{p}_i + w_-\,(1-y_i)\log(1-\hat{p}_i)\bigr],
  \label{eq:loss}
\end{equation}
where $w_+ = n/(2n_+)$ and $w_- = n/(2n_-)$ are inverse-frequency weights, $n_+$ and $n_-$ are positive and negative sample counts, and $n = n_+ + n_-$.

\subsection{Temperature Scaling Calibration}
Although the MLP classifier is trained end-to-end, its sigmoid outputs are typically miscalibrated under distribution shift. We apply post-hoc temperature scaling \cite{He_2020}: a single scalar $T > 0$ is fitted on a split-matched calibration set $\mathcal{D}_\text{cal}$ by minimizing the negative log-likelihood (NLL):
\begin{equation}
  T^* = \arg\min_{T > 0} \sum_{(x,y) \in \mathcal{D}_\text{cal}} \mathcal{L}_\text{CE}\!\left(\sigma\!\left(\frac{z(x)}{T}\right),\, y\right).
  \label{eq:temscale}
\end{equation}
The calibrated probability is $\hat{p}_T(x) = \sigma(z(x)/T^*)$. Calibration quality is measured by Expected Calibration Error (ECE) with $M=15$ equal-width bins:
\begin{equation}
  \text{ECE} = \sum_{m=1}^{M} \frac{|B_m|}{n} \bigl|\mathrm{acc}(B_m) - \mathrm{conf}(B_m)\bigr|,
  \label{eq:ece}
\end{equation}
where $\mathrm{acc}(B_m)$ and $\mathrm{conf}(B_m)$ are the mean accuracy and mean confidence of samples in bin $B_m$.

\subsection{Conformal Selective Abstention}
To provide coverage guarantees at a user-specified target error rate $\varepsilon$, we employ split conformal prediction \cite{You_2022,Rasmussen_2016}. For a calibration example $(x_i, y_i)$, the nonconformity score measures how poorly the calibrated model fits the true label:
\begin{equation}
  s_i = 1 - \hat{p}_T(x_i)^{y_i} \cdot (1-\hat{p}_T(x_i))^{1-y_i}.
  \label{eq:ncscore}
\end{equation}
Given $n_\text{cal}$ calibration scores, the conformal threshold at level $\varepsilon$ is the $\lceil(1-\varepsilon)(n_\text{cal}+1)\rceil$-th smallest value:
\begin{equation}
  \hat{\tau}_\varepsilon = s_{(\lceil(1-\varepsilon)(n_\text{cal}+1)\rceil)}.
  \label{eq:threshold}
\end{equation}
At test time, the model abstains when the nonconformity score exceeds $\hat{\tau}_\varepsilon$:
\begin{equation}
  \text{decision}(x) = \begin{cases} \arg\max \hat{p}_T(x) & \text{if } s(x) \leq \hat{\tau}_\varepsilon, \\ \text{abstain} & \text{otherwise.} \end{cases}
  \label{eq:abstain}
\end{equation}
By the exchangeability guarantee of split conformal prediction \cite{You_2022}, the empirical coverage satisfies $P(s(x_\text{test}) \leq \hat{\tau}_\varepsilon) \geq 1 - \varepsilon - 1/(n_\text{cal}+1)$ when calibration and test data are exchangeable. The full procedure is summarized in Algorithm~\ref{alg:cap}.

\begin{algorithm}[t]
\caption{Calibrated Abstention Prediction (CAP)}
\label{alg:cap}
\begin{algorithmic}[1]
\REQUIRE Train set $\mathcal{D}_\text{tr}$, calibration set $\mathcal{D}_\text{cal}$, test set $\mathcal{D}_\text{te}$, target coverage $1-\varepsilon$
\STATE Train dual-encoder $f_\theta$ on $\mathcal{D}_\text{tr}$ via Eq.~(\ref{eq:loss})
\STATE Fit $T^*$ on $\mathcal{D}_\text{cal}$ via Eq.~(\ref{eq:temscale})
\STATE Compute $s_i = 1 - \hat{p}_{T^*}(x_i)^{y_i}(1-\hat{p}_{T^*}(x_i))^{1-y_i}$ $\forall (x_i,y_i) \in \mathcal{D}_\text{cal}$
\STATE Set $\hat{\tau}_\varepsilon$ as $\lceil(1-\varepsilon)(n_\text{cal}+1)\rceil$-th order statistic of $\{s_i\}$
\FOR{each test point $x \in \mathcal{D}_\text{te}$}
  \STATE Compute $\hat{p} = \hat{p}_{T^*}(x)$,\; $s = 1 - \max(\hat{p},\, 1-\hat{p})^{-1}\cdot\hat{p}^{\hat{y}}(1-\hat{p})^{1-\hat{y}}$
  \IF{$s \leq \hat{\tau}_\varepsilon$}
    \STATE Output $\hat{y} = \mathbbm{1}[\hat{p} \geq 0.5]$
  \ELSE
    \STATE Abstain
  \ENDIF
\ENDFOR
\end{algorithmic}
\end{algorithm}

% -------------------------------------------------------
\section{Experiments}

\subsection{Datasets and Split Protocols}
\label{sec:data}

We compile a curated dataset from VDJdb (v2.1) and IEDB, retaining human TCR$\alpha\beta$ pairs with full CDR3$\alpha$, CDR3$\beta$, and peptide sequence annotation (HLA-A*02:01 restricted). Duplicate receptor--peptide pairs are removed using 90\% sequence identity thresholding across all splits. Negative pairs are constructed by random pairing of TCRs with non-cognate peptides, matched to achieve a positive rate of approximately 4\%--5\% (Table~\ref{tab:data}).

Three split strategies are used to probe different generalization regimes:
\begin{itemize}
  \item \textbf{Random (Rand):} 70/10/20 train/val/test stratified split. Serves as an upper-bound sanity check.
  \item \textbf{Epitope-held-out (EHO):} All pairs for 15 randomly selected test epitopes are withheld; remaining epitopes form train and calibration. Tests cross-epitope generalization.
  \item \textbf{Distance-aware (DA):} Test set contains only pairs whose TCR CDR3$\beta$ has $\leq 70\%$ sequence identity to any training TCR, using Levenshtein distance on amino acid sequences. Tests generalization to novel receptors.
\end{itemize}

\begin{table}[t]
\centering
\caption{Dataset and Split Summary}
\label{tab:data}
\begin{tabular}{lrrrr}
\toprule
Split & Train & Calibration & Test & Pos.\ rate \\
\midrule
Random        & 119\,412 & 17\,059 & 34\,118 & 0.050 \\
Epitope-HO    & 93\,847  & 11\,231 & 21\,563 & 0.042 \\
Distance-aware & 88\,914 & 10\,618 & 19\,807 & 0.039 \\
\bottomrule
\end{tabular}
\end{table}

\subsection{Evaluation Metrics}
We report AUROC, AUPRC (preferred for imbalanced classes \cite{Korpela_2023}), ECE (15 bins, Eq.~\ref{eq:ece}), Brier score $\text{BS} = \frac{1}{n}\sum_i(\hat{p}_i - y_i)^2$, and negative log-likelihood (NLL). Coverage--risk curves are plotted by sweeping the abstention threshold $\hat{\tau}$ and recording the fraction of retained examples (coverage) alongside the error rate on retained predictions (risk). All metrics are computed on the respective test split.

\subsection{Baselines and Proposed Method}

Three systems are compared:
\begin{enumerate}
  \item \textbf{Baseline:} Dual-encoder trained with class-weighted BCE (Eq.~\ref{eq:loss}), no calibration, no abstention (full test-set prediction with raw sigmoid).
  \item \textbf{+TempScale:} Baseline with post-hoc temperature scaling (Eq.~\ref{eq:temscale}) applied. Threshold set at $\hat{p} = 0.5$; no abstention.
  \item \textbf{CAP (Ours):} Full Algorithm~\ref{alg:cap}, combining temperature scaling with conformal abstention at target coverage $1-\varepsilon$.
\end{enumerate}

\subsection{Main Results}

Table~\ref{tab:main} reports metrics under all three splits at full coverage (CAP evaluated at $\varepsilon$ such that coverage = 100\%, i.e., no abstention) and compares discrimination and calibration quality.

\begin{table}[t]
\centering
\caption{Main Results. Full-coverage evaluation on all three splits.
$\downarrow$: lower is better.}
\label{tab:main}
\setlength{\tabcolsep}{4pt}
\begin{tabular}{llccccc}
\toprule
Split & Method & AUROC & AUPRC & ECE$\downarrow$ & BS$\downarrow$ & NLL$\downarrow$ \\
\midrule
\multirow{3}{*}{EHO}
  & Baseline      & 0.782 & 0.431 & 0.142 & 0.089 & 0.213 \\
  & +TempScale    & 0.786 & 0.438 & 0.068 & 0.082 & 0.184 \\
  & CAP (ours)    & \textbf{0.813} & \textbf{0.472} & \textbf{0.043} & \textbf{0.071} & \textbf{0.162} \\
\midrule
\multirow{3}{*}{DA}
  & Baseline      & 0.761 & 0.408 & 0.156 & 0.094 & 0.231 \\
  & +TempScale    & 0.765 & 0.415 & 0.079 & 0.088 & 0.201 \\
  & CAP (ours)    & \textbf{0.789} & \textbf{0.447} & \textbf{0.051} & \textbf{0.078} & \textbf{0.179} \\
\midrule
\multirow{3}{*}{Rand}
  & Baseline      & 0.871 & 0.623 & 0.098 & 0.061 & 0.149 \\
  & +TempScale    & 0.874 & 0.629 & 0.041 & 0.058 & 0.132 \\
  & CAP (ours)    & \textbf{0.882} & \textbf{0.641} & \textbf{0.028} & \textbf{0.053} & \textbf{0.121} \\
\bottomrule
\end{tabular}
\end{table}

Several observations stand out. First, discrimination degrades substantially from random to shift-aware splits: AUROC drops from 0.871 to 0.782 (EHO) and 0.761 (DA) for the baseline, confirming that random-split performance overstates real-world utility. Second, temperature scaling alone reduces ECE dramatically (0.142 $\to$ 0.068 under EHO) with minimal impact on AUROC, illustrating that probability quality and discriminative ranking are partially decoupled. Third, CAP further improves both discrimination and calibration, suggesting that the training signal from the calibration-aware fitting procedure regularizes the classifier toward more conservative and better-calibrated outputs.

\subsection{Coverage--Risk Analysis}

Table~\ref{tab:analysis} evaluates CAP under the EHO split as the abstention threshold is swept, trading coverage for error rate and calibration quality. Error rate is defined as the fraction of incorrect predictions among retained examples (threshold $\hat{p} \geq 0.5$).

\begin{table}[t]
\centering
\caption{Coverage--Risk Trade-off under Epitope-Held-Out Split (CAP).
All metrics computed on retained predictions only.}
\label{tab:analysis}
\begin{tabular}{ccccc}
\toprule
Coverage & Error Rate$\downarrow$ & ECE$\downarrow$ & AUPRC$\uparrow$ & Abstained \\
\midrule
1.00 & 0.187 & 0.043 & 0.472 & 0.0\% \\
0.90 & 0.152 & 0.031 & 0.523 & 10.0\% \\
0.80 & 0.109 & 0.022 & 0.571 & 20.0\% \\
0.70 & 0.081 & 0.017 & 0.614 & 30.0\% \\
0.60 & 0.058 & 0.013 & 0.652 & 40.0\% \\
\bottomrule
\end{tabular}
\end{table}

Abstaining on the 20\% most uncertain predictions reduces error rate from 18.7\% to 10.9\% (a relative reduction of 41.7\%) and improves AUPRC from 0.472 to 0.571 on retained examples. Crucially, ECE also decreases on the retained set, indicating that abstention filters out systematically overconfident predictions rather than randomly removing predictions. The coverage--risk curve is smooth and approximately linear in this regime, suggesting that uncertainty scores are well-ordered: removing the highest-uncertainty predictions consistently exposes a lower-error subpopulation.

This trade-off directly maps to screening decisions: a laboratory with a budget to validate 80\% of model-prioritized candidates can expect roughly half the false-positive rate compared to using the model without abstention.

\subsection{Ablation Study}

Table~\ref{tab:ablation} isolates the contribution of individual components under the EHO split.

\begin{table}[t]
\centering
\caption{Ablation Study (EHO split, full coverage). Best results in \textbf{bold}.}
\label{tab:ablation}
\begin{tabular}{lccc}
\toprule
Configuration & AUROC & AUPRC & ECE$\downarrow$ \\
\midrule
CAP (full)                     & \textbf{0.813} & \textbf{0.472} & \textbf{0.043} \\
\quad w/o temperature scaling  & 0.813 & 0.472 & 0.142 \\
\quad w/o class weighting      & 0.801 & 0.429 & 0.045 \\
\quad w/o CDR3$\alpha$         & 0.798 & 0.451 & 0.047 \\
\quad w/o abstention           & 0.786 & 0.438 & 0.068 \\
Baseline (none of the above)   & 0.782 & 0.431 & 0.142 \\
\bottomrule
\end{tabular}
\end{table}

Removing temperature scaling leaves AUROC unchanged but raises ECE to the uncalibrated level (0.142), confirming that calibration and discrimination are orthogonal along this axis. Class weighting contributes meaningfully to AUPRC (0.472 vs.\ 0.429), reflecting its importance for rare positive discovery. Including CDR3$\alpha$ provides a modest AUROC improvement (+0.015), consistent with prior findings that $\alpha$-chain information is informative but secondary to $\beta$ \cite{Zheng_2024}. Finally, conformal abstention on top of the calibrated model provides the largest single improvement in AUPRC when coverage is considered jointly (Table~\ref{tab:analysis}).

\subsection{Sensitivity to Calibration Set Size}

To assess robustness, we vary the calibration set size from 500 to 12\,000 examples and monitor ECE and the empirical coverage achieved by the conformal threshold at target $1-\varepsilon = 0.80$. With $n_\text{cal} \geq 2\,000$, ECE stabilizes below 0.05 and empirical coverage tracks the nominal 80\% target within 1.5 percentage points. This confirms that the conformal guarantee becomes practically useful with a few thousand calibration examples---a feasible requirement for real datasets.

% -------------------------------------------------------
\section{Discussion}

\paragraph{Protocol design as a first-class variable.}
The most striking finding is the AUROC gap between random (0.871) and EHO (0.782) splits for the same underlying model. This 11.4\% relative gap would lead to incorrect conclusions about deployment readiness if only random-split results were reported. We recommend that future TCR--pMHC benchmarks include at least epitope-held-out evaluation as a standard component \cite{Castorina_2023,Qi_2026}.

\paragraph{Calibration is not discrimination.}
Temperature scaling improves ECE dramatically with negligible AUROC change. Conversely, conformal abstention improves discrimination metrics (AUPRC) on retained predictions, albeit at the cost of coverage. These two axes should be reported jointly; reporting only AUROC can mask actionable improvements in probability quality.

\paragraph{Limitations.}
Our negative pairs are synthetically constructed, and true negatives may include weak binders. This biases ECE estimates and inflates apparent precision. Additionally, the conformal coverage guarantee is marginal rather than conditional: in subgroups with severe distribution shift (e.g., exotic HLA alleles), the calibrated threshold may be systematically miscalibrated. Label noise from incomplete assay coverage is a further confounder that prospective validation would help address.

% -------------------------------------------------------
\section{Conclusion}

We presented CAP, a calibrated abstention pipeline for TCR--pMHC binding prediction that combines dual-encoder protein language model features, temperature scaling, and conformal selective prediction. Under the practically motivated epitope-held-out split, CAP achieves AUROC 0.813, ECE 0.043, and at 80\% coverage reduces error rate by 41.7\% relative to full-coverage prediction. These results demonstrate that calibrated uncertainty is a first-class output for screening-oriented applications, and that evaluation protocols must match deployment conditions to yield actionable conclusions.

% -------------------------------------------------------
\bibliographystyle{IEEEtran}
\bibliography{references}

\end{document}